# Preliminary design of the Visible Spectro-Polarimeter for the Advanced Technology Solar Telescope

Alfred G. de Wijn[*,a], Roberto Casini[a], Peter G. Nelson[b], and Pei Huang[c]

[a]High Altitude Observatory, National Center for Atmospheric Research, P.O. Box 3000, Boulder, CO 80303, USA; [b]Sierra Scientific Solutions, 1540 Patton Drive, Boulder, CO, USA; [c]Consultant to NCAR, 1484 N. Larkspur Ct., Lafayette, CO, USA

## ABSTRACT

The Visible Spectro-Polarimeter (ViSP) is one of the first light instruments for the Advanced Technology Solar Telescope (ATST). It is an echelle spectrograph designed to measure three different regions of the solar spectrum in three separate focal planes simultaneously between 380 and 900 nm. It will use the polarimetric capabilities of the ATST to measure the full Stokes parameters across the line profiles. By measuring the polarization in magnetically sensitive spectral lines the magnetic field vector as a function of height in the solar atmosphere can be obtained, along with the associated variation of the thermodynamic properties. The ViSP will have a spatial resolution of 0.04 arcsec over a 2 arcmin field of view (at 600 nm). The minimum spectral resolving power for all the focal planes is 180,000. The spectrograph supports up to 4 diffraction gratings and is fully automated to allow for rapid reconfiguration.

## 1. INTRODUCTION

Our understanding of solar magnetism depends on our ability to detect and interpret the polarization signatures of magnetic fields in solar spectral lines. Since the identification of sunspots as regions of strong magnetism by the observation of the Zeeman effect in 1908,[1] the field of solar physics has become increasingly dependent on measurements of magnetic field in the solar atmosphere. Photospheric vector magnetic field measurements are now routinely made with various instruments. Several other magneto- and electro-optical effects are now also starting to be used as diagnostics of solar plasma parameters. While the Zeeman effect is mostly blind to weak fields, the Hanle effect observed in linear polarization has been exploited to diagnose weak turbulent magnetic fields[2,3,4] and the magnetic structure of solar prominences.[5] A rich spectrum of linear polarization signatures observed near the solar limb (the "second solar spectrum"[6]) is actively being investigated to constrain solar magnetic fields and improve our understanding of light-scattering physics. Observations of lines that display hyperfine structure like those of Mn I have been used to break the intrinsic degeneracy between magnetic flux and magnetic field that arises when using the Zeeman effect for weak magnetic fields.[7,8,9] Information about plasma kinetics during solar flares is encoded in the linear polarization of H lines through impact polarization.[10]

Because of the complexity of the physics involved in polarized radiative transfer and line formation as well as the intricate structure of the solar atmosphere where these polarization signatures are produced, these studies profit highly from the simultaneous observations of different spectral lines. For example, the application of line-ratio techniques to multiple spectral lines has been shown to greatly enhance the diagnostic potential of Zeeman-effect observations.[11] By probing different heights in the solar atmosphere, observing multiple lines also serves to constrain the magnetic field geometry in three dimensions. Improvements in detector technology have also made it possible to take advantage of the increased sensitivity of the Zeeman effect in the infrared.[12,13]

The ATST's Visible Spectro-Polarimeter (ViSP) is an echelle spectrograph designed to provide precision measurements of the full state of polarization across spectral line profiles. Its objective is to measure three highly resolved spectra simultaneously at diverse wavelengths from 380 to 900 nm by imaging three spectra onto different focal planes (cameras). The ViSP is also distinct from many spectrographs in that it seeks to obtain this information over a wide field of view (2 arcmin) with an exceptionally high spatial resolution (half the diffraction limit of the ATST, or approximately 0.04 arc seconds at 600 nm). This nearly 4,000:1 spatial dynamic range is unprecedented for solar spectrographs.

---

[*] dwijn@ucar.edu, phone +1 303 497 2171

Not all science requires polarimetric information. Non-polarimetric observations are important because the time to build up a map of a region can be dramatically reduced if no magnetic information about that region is required. Temperature and velocity fields can be obtained from spectral information alone. The ViSP can be run in a non-polarimetric mode with no modifications. In such a mode the scan rate is limited by the camera speed and the desired spatial resolution. The spatial resolution is given by the slit velocity in arcsec/s divided by the camera frame rate.

The ViSP is designed to be a fully automated spectrograph capable of observing any triplet of lines from 380 to 900 nm with optical efficiencies of 30% or higher. A computer-controlled interface allows configuration of the spectrograph with little or no manual intervention. The ability to rapidly reconfigure the spectrograph makes it ideal for multiple observation campaigns within the same day or to address targets of opportunity (such as solar flares). Key to the success of the ViSP's mission is to obtain a spectral resolution of 180,000 or greater from 380 to 900 nm while maintaining good polarimetric sensitivity for any triplet of lines desired by the campaign. To achieve this goal, and to ensure high optical efficiency across the spectrum the spectrograph must be equipped with a variety of gratings selectable by an automated grating turret. For high-precision polarimetry the spectrograph must also employ 'dual-beam' configuration where two identical spectra but orthogonal in the direction of polarization analysis are formed at each of the three focal planes. This combined with rapid polarization modulation and fast camera readout will enable the ViSP to obtain photon-noise limited polarimetry.

## 2. GENERAL CHARACTERISTICS

The requirements of the ViSP instrument imposed by the target science are summarized in Table 1. We discuss several more critical requirements that drive the design in detail below.

The spectral coverage of the ViSP should include molecular bands and the Ca II H and K lines in the range of 380–400 nm. At longer wavelengths it is important to cover the Ca II infrared lines near 850 nm and other lines between 850 and 900 nm that have been used for polarimetry. While the ViSP will be compatible with measuring the He I 1083 nm line and beyond, its ability to do so will be limited by the quantum efficiency of visible-light detectors. If equipped with IR detectors the ViSP will be compatible with wavelengths up to 1.6 microns. To benefit from multi-line diagnostics the ViSP should be capable of observation in three different wavelength bands simultaneously.

Polarimetric sensitivity is ultimately limited by photon statistics. To meet polarimetric sensitivity requirements the ViSP must have a sufficiently high optical efficiency. Because efficiency is a strong function of wavelength and spectrograph configuration (and in particular the angles of incident and diffracted light from the grating), the polarimetric sensitivity requirement can only be addressed by assuming specific ViSP configurations. If an observer chooses an 'unfortunate' combination of lines (for example) the efficiencies for all lines could be below nominal requirements to meet a specific type of observation or time cadence. In short, with a general-purpose instrument like the ViSP it is impossible to guarantee that a given polarimetric efficiency will be achieved in all configurations.

Table 1. General characteristics of the ViSP spectrograph.

| Parameter | Requirement | Comment |
|---|---|---|
| Wavelength range | 380–900nm | 380–1600 nm goal can be met |
| Simultaneous wavelengths | 3 focal planes | |
| Spatial resolution | 2× telescope resolution | Approx. 0.04 arcsec |
| Spatial field of view | 2 arcmin square | |
| Resolving Power | 180,000 at 630 nm | |
| Slit Scan Repeatability | +/– ½ a slit width | |
| Slit Scan Accuracy | 0.1 arcsec | |
| Polarimetric Sensitivity | $10^{-4}$ $I_c$ | |
| Polarimetric Accuracy | $5\times10^{-4}$ $I_c$ | Flows down to calibration requirements |
| Temporal Resolution | $10^{-3}$ $I_c$ in 10 seconds | |
| Spectral Bandpass | 1.1 nm at 630 nm | For broad spectral lines; coupled to FP size. |
| Setup time | At least 1 focal plane in 10 min | Should obtain alignment of all three focal planes in ~30 s |
| Slit move time | 200 ms for neighboring slit positions | Applies only to the polarimetric mode of the ViSP |
| Slit slew velocity | 2 arcmin in 30 s | |

Polarimetric accuracy is determined by how well various elements of the telescope-instrument system can be calibrated. The ViSP will be provided with polarization analysis and calibration software adequate to the scope of accurately measuring the Mueller matrix of the optical system following the calibration optics. The fulfillment of the polarimetric accuracy requirement depends also on the ability to properly measure the telescope matrix that precedes the calibration optics and must be determined by the facility.

The ViSP should have a field of view (FOV) capable of capturing the full extent of a moderately-sized active region. This will allow the full mapping of the active region's magnetic network that is necessary to model and diagnose structures that are the precursors to flares and coronal mass ejections. For this reason, the FOV should be at least 2 arcmin square. This sets the ratio of the FOV to the angular width of the entrance slit around 4000:1. Large format cameras will consequently be required to simultaneously fulfill the FOV and spatial/spectral resolution requirements.

The ATST will make extensive use of Adaptive Optics (AO) technology for science observations. To fully exploit this, the ViSP should have adequate optical imaging capabilities to not degrade the image quality arriving at the entrance slit of the spectrograph. Narrow spectrograph slits, however, greatly reduce the scan rate possible for sampling the field (scan time for a given FOV increases like [slit width]$^{-2}$). Experience at other telescopes has shown that polarimetry right at the diffraction limit is rarely very productive because of the strongly reduced MTF of the telescope and the limited signal-to-noise ratio. The ViSP entrance slit will therefore sample the field at twice the sample width required to reach the diffraction-limited performance of the ATST.

The ViSP is designed to reach a nominal spectral resolving power of 180,000 at 630 nm. The spectrograph resolving power is defined as the wavelength of light divided by $\Delta\lambda$, where $\Delta\lambda$ is the smallest difference in wavelength between two monochromatic waves that can be separated by the spectrograph. Because of the ViSP's large field of view (roughly a 4000:1 ratio of slit height to width) the image of the slit on the focal plane must be equal to a single pixel. Normally spectrographs try to 'critically sample' the image of the slit but doing so in ViSP would require cameras of at least 8,000 pixels in the slit direction (and such cameras are currently not expected to be available). As a consequence, for the ViSP $\Delta\lambda$ is defined as twice the FWHM of the image of the slit on the focal plane of the camera in units of wavelength. In designing the ViSP towards this goal, we took into account diffraction effects of the instrument's optics.

The resolving power of a spectrograph scales with wavelength only in a fixed spectrograph configuration and if the entrance slit limits the resolution. The ViSP is not slit-dominated in its resolution in most configurations. Resolution is a strong function of spectrograph geometry and how the spectrograph is configured to meet the multiple-wavelength requirement. The current spectrograph concept allows for the arms to move such that the angular difference between the light incident on the grating ($\alpha$) and the light diffracted from the grating ($\beta$) can be as large as 24 deg to achieve wavelength diversity. For $\beta < \alpha$ angles the spectral resolution of a spectrograph for a given configuration of the grating drops because of the effect of anamorphic magnification. The ViSP design mitigates this loss of resolving power by adopting camera lenses with constant f/# and diameters that increase with decreasing $\beta$ so to accommodate the larger widths of the diffracted beams. This can be desirable for lines with profiles in excess of 1 nm width.

## 3. PRELIMINARY OPTICAL DESIGN

### 3.1 Feed Optics

The main light feed to the coudé room is directed to the deformable mirror (DM) on which an image of the aperture is formed (a pupil). The light reflected from this passes through three beam splitters that split light off to different instruments. The light from the DM is collimated. After the last beam splitter the light passes to the ViSP feed optics table that holds the optics that image the field onto the slit. The light feeding the spectrograph is designed to be telecentric (the central rays for all points in the field are parallel) to keep the size of the beams and of optics from diverging over the optical path. This is achieved by placing the DM one focal length away from the imaging optics – thus the DM to slit distance is constrained and the feed optics table location is fixed.

The feed optics consist of a flat mirror that directs the light into the first element of a Dall-Kirkham mirror pair. These two mirrors form a diffraction-limited image onto the slit with a 120-arcsec square field of view (FOV). The image is approximately 81 mm square and is tilted by ~1.8 deg. This allows the light reflected from the slit jaws to be used in a context imaging system. A concept context imager design is shown in Figure 1 but is currently not planned for deployment at first light. The feed optics (less the slit) will be mounted on an optical table rigidly fastened to the main optical table of the ViSP. In this way the precise alignment requirements for the Dall-Kirkham pair are maintained.

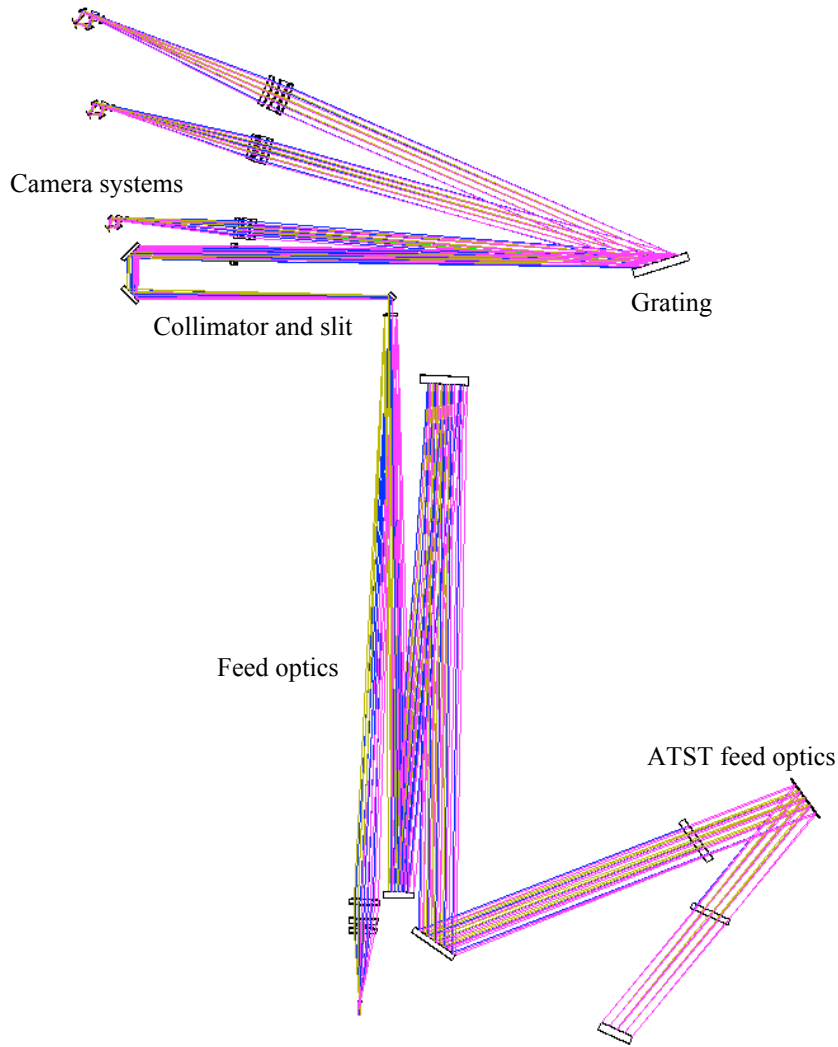

Figure 1. An overview of the ViSP spectrograph preliminary optical design in the ATST coudé room.

The concept context imaging system consists of a custom triplet lens, a filter wheel, and a camera package with automated focus and centering. The context camera must be re-focused for each filter. The context imager is able to locate the slit over the entire wavelength range of the ViSP and over most of the FOV.

### 3.2 Slit and Collimator

Figure 2 shows a detailed view of the slit, collimator doublet lens, and grating. The entrance slit and the first fold mirror that folds the light to the left are mounted on a precision linear stage. This stage is used to scan the field by moving the slit in the plane of the image. Immediately behind the entrance slit is a prism that is used to correct for the tilt in the focal plane when the spectrograph is used in a multiple-slit configuration. This prism allows the collimator to be a simple on-axis doublet design while adding a negligible amount of dispersion. Note that in the figure there are three beams emanating from the slit assembly. These represent three beams of a multi-slit configuration.

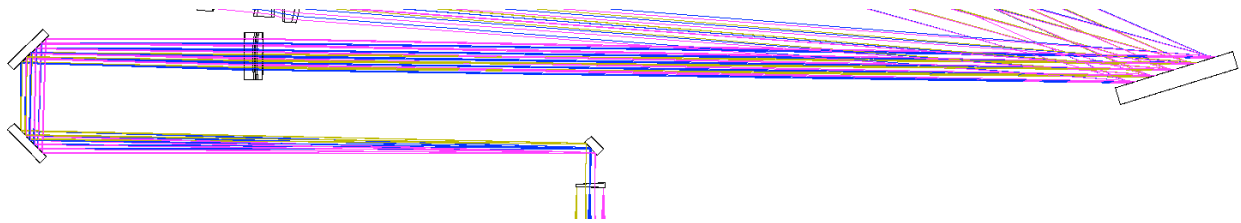

Figure 2. The slit, collimator and grating. Three beams emanating from the slit assembly represent an optional multi-slit configuration. The two fold mirrors at the left are scanned at half the rate of the slit to compensate the length between the slit and the collimator doublet lens. The grating on the right is on a turret that holds up to 4 gratings.

Because of the extreme precision requirements on the slit, it is fabricated using photo-lithography. To change the slit requires manual intervention in the coudé room. At first light there will be a library of three slit widths (optimized for the blue, red, and near-IR). The choice of an etched slit was also made necessary by the possibility of using a multi-slit configuration. The slit will be automated with clocking for alignment with other instruments but will not have decking (a FOV limiter). Alignment hairlines will be incorporated into the slit design.

The next set of optics is a pair of fold mirrors mounted on a linear stage. By moving this stage at half the rate of the entrance slit the distance from the slit to the collimating lens is maintained and thus the collimation of the light is preserved throughout the scan. Note that as a pair these mirrors correct for small wobble in the stage's motion and therefore the specifications on this stage are much less stringent than the entrance slit stage.

The focal length of the collimating lens is 2.6 m. It is an achromatic doublet with only spherical surfaces. The grating is placed 2.6 m from the lens, producing an image of the telescope pupil on the grating. This minimizes the size of the grating because of the waist in the beam at that position and also ensures that all points along the slit illuminate the same area of the grating. The grating turret is a vertical stack (out of the page; it passes through a hole in the optical table) of up to 4 gratings (2 will be provided at first light). An elevator motion selects the grating to be used and a high-precision rotational stage is used to set the α-angle of the incident light. Each grating is on a removable cassette for ease of replacement.

### 3.3 Camera Assemblies

Figure 3 shows one of the three camera assemblies. All the components in this figure are part of a single mechanical package that is moved in an arc centered on the diffraction grating. The motion is such that the package is always pointed towards the grating. This motion allows for different diffraction β angles to be selected. The three camera packages are designed with a common f/7 focal ratio. However, each camera package has a different aperture because of anamorphic magnification: at smaller β angles the beam diameter increases (this is easily seen in Figure 1). Anamorphic magnification also has the effect of making the image of the slit narrower at the camera, which means that the focal lengths must increase to match the slit width to pixel size and thus preserve the spectrograph's resolving power.

There is a trade associated with the width of the camera aperture. To have a non-vignetted spectrum the aperture needs to be large enough to accommodate all the β angles associated with the spectral width on the focal plane. Doing so, however, makes the camera packages very wide and thus increases the minimum Δβ between two beams to avoid mechanical interference with adjacent cameras. Minimizing Δβ between beams is a key factor for wavelength diversity in the spectrograph. To decrease Δβ we therefore opted to allow some vignetting near the edges of the spectrum. The impact of this is expected to be minimal to the instrument's mission.

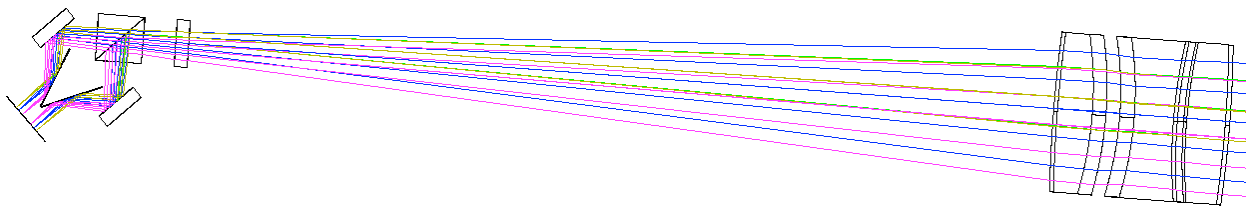

Figure 3. One of the three camera assemblies. Each assembly has a different focal length but a common f/7 focal ratio.

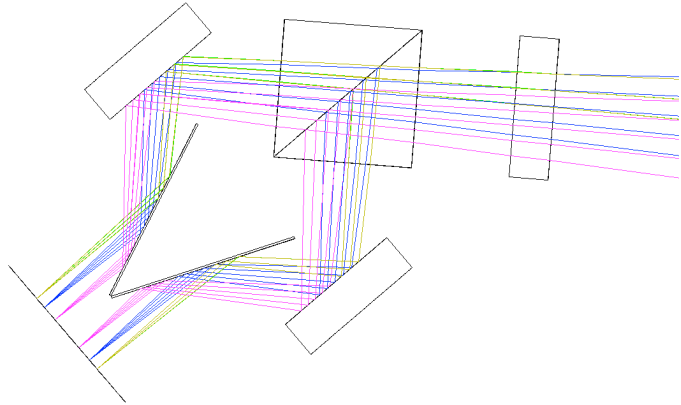

Figure 4. Detail of the polarization analyzer. The light enters from the right. A polarizing beam splitter divides the beam and two mirrors reflect the beams onto a reflecting prism that make the two beams adjacent and parallel. One beam experiences three reflections and the other two, producing two mirror-image spectra.

The first element in the camera system is the triplet lens. The glasses and design of the lens for each package is very similar except for the focal length and aperture. A triplet lens design is needed mainly to correct aberrations in the field and not chromatic aberrations. Because each arm is essentially monochromatic, the lens optimization was allowed to use a different focal length over the spectrum. This means the cameras will need re-focusing when moving to different spectral regions. This minor inconvenience helps to make the lens design feasible and cost effective.

Two strategies exist for the order isolation filters. The first is to split the spectral range into regions equal to ~½ the free spectral range of the gratings. For gratings working in orders 5–15 this would require ~20 filters. This universal approach is balanced by the problem that spectral lines near the border between two filters may have less than optimal efficiency. The second strategy is to have broad filters centered on spectral lines of interest. This optimizes filter performance but may require more filters in the long run. The strategy we assume is a hybrid of the two: a cassette with 30 positions will accommodate 'general purpose' isolation filters and a limited number of line-centered filters for high-priority spectral lines. The cassette design is similar to a slide projector. A long tray will hold a stack of filters in rectangular mounts. The tray will index so the desired filter is above the filter location and an arm will push the filter from the tray into the optical path. This tray concept is adaptable to accept many more filters if required.

The polarizing beam splitter is one of the most challenging aspects of the ViSP design. The device must introduce very low wavefront error and have good transmission and reflection efficiencies over the entire spectral operating range of ViSP (380–900 nm). This capability has been demonstrated using conventional thin-film technology.

Figure 4 shows a detailed view of the order-isolating filter, polarizing beam splitter, and focal plane. After the polarizer is a pair of flat mirrors that direct the beams towards a prism. These mirrors will have tip, tilt, and piston adjustments to allow the planes of focus for the two beams to be made parallel and coplanar. Their adjustment is manual and only performed once. Finally there is a 45 deg prism that directs the beams to the focal plane. This prism and the polarizing beam splitter are fixed, making them fiducial points of the system. The difference in reflectivity between P- and S-polarizations for this prism is a possible issue because of the 67.5 deg angle of incidence. Modeling of protected silver coatings shows that this difference is minor (a few percent) and typically negligible when compared to the polarizing properties of the diffraction grating.

## 4. HARDWARE OVERVIEW

The ViSP mechanical system provides the ability to configure the instrument without coudé room access through the use of precision automated actuators and high mechanical stability. The mechanisms feature low systemic deviations and offer the precision and long-term observational stability required.

The ViSP experiment is mounted to a group of joined optical tables on rigid legs coupled to the coudé room sub-floor. Figure 5 shows a bird's eye view of the ViSP instrument in the ATST coudé room. The components are grouped into two areas. The feed optics table holds the steering and powered optics required to take the ATST collimated input beam and form an image on the spectrograph's entrance slit (light green). The main optic table carries the entrance slit and all post-slit optics and focal plane systems.

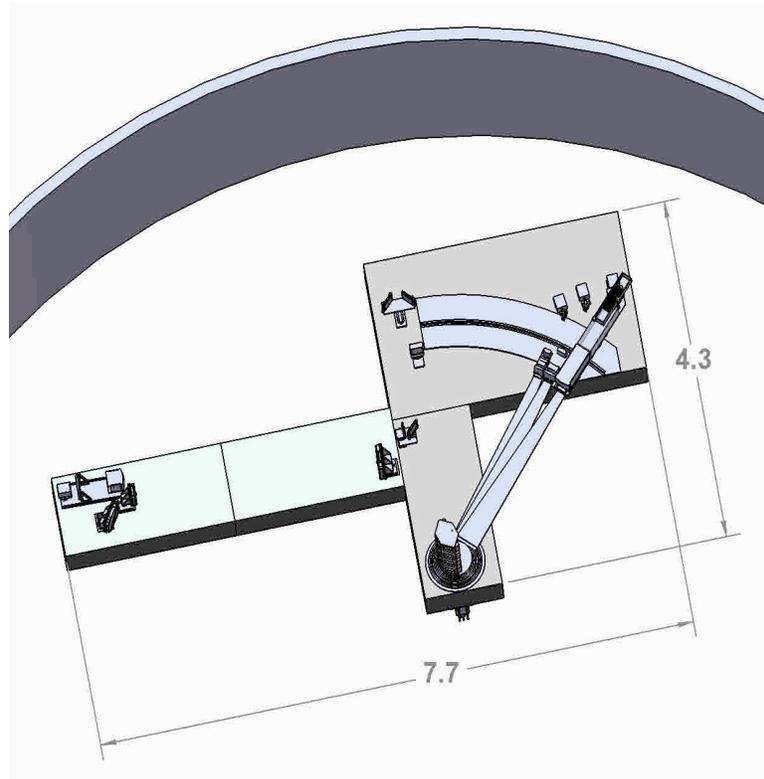

Figure 5. Overview of the ViSP instrument in the ATST coudé room. Dimensions are given in meters. The ATST feed optics are not shown.

## ACKNOWLEDGEMENTS

The National Center for Atmospheric Research is sponsored by the National Science Foundation.